\documentclass[english,aps,prl,twocolumn,floatfix,superscriptaddress,a4paper,showkeys,nofootinbib]{revtex4-2}
\usepackage[colorlinks=true,linktocpage=true,linkcolor=blue,citecolor=blue,allcolors=blue]{hyperref}
\usepackage[english]{babel}
\usepackage{graphicx}
\graphicspath{ {figures/} }
\usepackage{latexsym}
\usepackage[utf8]{inputenc}
\usepackage{xspace}
\usepackage{indentfirst}
\usepackage{enumitem}
\usepackage{xcolor}
\usepackage{soul}
\usepackage{amsmath}
\usepackage{amssymb}
\usepackage{url}

\newcommand{\mean}[1]{\langle #1 \rangle}

\newcommand{\subeq}[1]{\begin{subequations}\begin{align} #1 \end{align}\end{subequations}}

\usepackage[normalem]{ulem}
\usepackage{xcolor}

\begin{document}

\title{
Freeze-out model of light nuclei formation in heavy-ion collision transport
}

\author{Oleh Savchuk}\thanks{Corresponding author}
\email{savchuk@frib.msu.edu}
\affiliation{International Research Laboratory on Nuclear Physics and Astrophysics, Michigan State University and CNRS, East Lansing, MI 48824, USA}

\author{Pawel Danielewicz}
\affiliation{Department of Physics and Astronomy and the Facility for Rare Isotope Beams, Michigan State University, East Lansing, MI 48824 USA}

\author{William Lynch}
\affiliation{Department of Physics and Astronomy and the Facility for Rare Isotope Beams, Michigan State University, East Lansing, MI 48824 USA}

\author{J\'er\^ome Margueron}
%ORCID: 0000-0001-8743-3092 
\affiliation{International Research Laboratory on Nuclear Physics and Astrophysics, Michigan State University and CNRS, East Lansing, MI 48824, USA}
\date{\today}

\begin{abstract}
Cluster production plays an important role in heavy-ion collisions at intermediate beam energies, where light nuclei contribute substantially to final-state yields and to other observables that are used to infer the nuclear equation of state. 
In this letter, we propose a new approach for clustering that combines dynamical transport and thermal cluster production for mid-rapidity particles.
The resulting hybrid coarse-graining model matches nucleon and light-cluster descriptions at freeze-out while properly accounting for thermal non-uniformity and collective transport in the hot, strongly interacting systems created in heavy-ion collisions. To illustrate the capabilities of this model, yields at 4~fm impact parameter, spectra and elliptic flows at 7.4~fm ($20\text{--}30\%$ centrality) are predicted at mid-rapidity for semi-peripheral Au$+$Au collisions at an incident energy of $E_\text{lab}=1.23~A\mathrm{GeV}$.
\end{abstract}

\maketitle

Clusters are produced in abundance across a broad spectrum of collision energies, at facilities such as FRIB, RIBF-RIKEN, FAIR,  GANIL, RHIC, and LHC~\cite{Kurata-Nishimura:2025ydt,HADES:2022gms,FOPI:1996cjz,FOPI:2010xrt,ALICE:2015wav}. These light clusters are seemingly fragile probes of the medium as they are immersed in what is essentially a nuclear hellfire~\cite{Siemens:1979dz} that should vaporize them.
The nature of their persistent appearance has, therefore, been a subject of discussion within the community~\cite{Braun_Munzinger_2019}. It is admitted that a deeper understanding of cluster formation and survival is essential in Heavy-Ion Collision (HIC), since these particles serve as sensitive probes into the properties and evolution of strongly interacting matter.

One of the main goals of HIC studies is to infer the high-density nuclear Equation of State (EoS).  According to transport simulations of collisions, the most promising observables for assessing the pressures generated in the central high-density region are those that quantify transverse anisotropies of the collective motion, such as sideward and elliptic flows~\cite{Brachmann:1999xt,Rischke:1995pe,Steinheimer:2022gqb,Stoecker:2004qu}. Among the products of collisions, clusters are more sensitive to collective motion than nucleons~\cite{Kurata-Nishimura:2025ydt,HADES:2022gms,FOPI:1996cjz,FOPI:2010xrt,Danielewicz:1999vh}, so clusters emerging from the central region could provide the best observables to constrain the nuclear EoS.

Currently, cluster production is harder to predict than nucleons in transport models~\cite {PhysRevC.21.1301,Kireyeu:2025jrb,SATO1981153,PhysRevLett.37.667,MROWCZYNSKI199243,CSERNAI1986223,Liu_2019,SUN2018499,Polleri_1998}. In only a few models~\cite{Oliinychenko:2020znl,Staudenmaier:2021lrg,DANIELEWICZ1991712,Kanada-Enyo:2012yif} are clusters produced dynamically and treated as independent particles. In most models, clusters are, however, produced by postprocessing transport results, with the most prominent methods coalescence and simulated annealing~\cite{Botvina:2020yfw,PhysRevC.46.2002,PhysRevC.63.034605,Bratkovskaya:2025oys,Coci:2024aue}. 

Here, we propose a novel approach to the production of light clusters with transport. 
The approach exploits several HIC observations: 
\begin{enumerate}
\item In transport simulations of central HIC, local thermal equilibrium is reached at late stages of evolution~\cite{Bravina:1998pi,Bravina:1999dh, danielewiczDeterminationEquationState2002,Inghirami:2019muf}. 
\item In simulations with dynamic production, the final clusters emerge during the thinning of nuclear matter~\cite{PhysRevC.51.716}.  
\item Global chemical equilibrium models can well describe measured yields of these light clusters at different collision energies~\cite{Letessier:2005qe,Andronic:2017pug}.
\end{enumerate}

To circumvent technical difficulties in dynamic cluster-production modeling, whose details may be questioned, we propose a simple description in which cluster production occurs instantaneously in a single chemical freeze-out step~\cite{Huovinen:2012is}, with yields and momentum distributions consistent with local equilibrium~\cite{landau1987fluid} in the presence of a collective flow field.  The state after formation may either be considered final, or transport may be followed without inelastic processes until kinetic freeze-out is reached.  Notably, the freeze-out concept is often successfully employed in descriptions of ultrarelativistic collisions while making the transition from hydrodynamic to transport descriptions~\cite{PhysRevC.60.021902}. 
Here we apply it to the state resulting from a large number of collisions, see hereafter.

This chemical freeze-out approximation has the virtue that it enables simple predictions for the production of excited states of clusters, such as alpha particles.  In ultrarelativistic collisions, a freeze-out approximation can also be applied to the production of pions and baryon resonances.  In intermediate-energy collisions, the chemical freeze-out for those particles likely occurs earlier. In our transport description, we therefore produce pions and baryon resonances earlier in transport, not during the later freeze-out of the clusters.

For cluster production, we apply a simple freeze-out condition during the expanding phase of the participant matter, which is related to the local density. Specifically, participant nucleons propagate as nucleons until they reach the low freeze-out density: $n_\text{fr}$. Then, the cluster and free-nucleon yields are calculated assuming chemical freeze-out.  Afterward, the resulting clusters and free nucleons propagate to the detectors without further reactions. Spectator matter that does not reach $n_\text{fr}$ in the transport model does not directly disintegrate. Its decay
% during the transport phase of the description
 can be addressed via statistical decay models~\cite{Bondorf:1995ua}.
 
For a given incident energy and impact parameter, we define the matching conditions by equating the densities of conserved quantities $X_\text{tr}$ at $n_\text{fr}$ to their local equilibrium values, assuming a gas composed of both nucleons and light clusters $X_\text{INCG}$ (see Supplemental Material), where $X$ is the energy density $\varepsilon$, the baryon density $n_B$, the isospin density $n_{I_3}$, and the strangeness density $n_S$:
\subeq{
\mean{\varepsilon_\text{tr}}(r_\text{fr},t_\text{fr}) &= \varepsilon_\text{INCG}[T, \mu_B, \mu_{I_3}, \mu_S](r_\text{fr},t_\text{fr}), \label{eq:energy}\\
\mean{n_{B,\text{tr}}}(r_\text{fr},t_\text{fr}) &= n_{B,\text{INCG}}[T, \mu_B, \mu_{I_3}, \mu_S](r_\text{fr},t_\text{fr}), \label{eq:nb}\\
\mean{n_{I_3,\text{tr}}}(r_\text{fr},t_\text{fr}) &= n_{I_3,\text{INCG}}[T, \mu_B, \mu_{I_3}, \mu_S](r_\text{fr},t_\text{fr}), \label{eq:isospin}\\
\mean{n_{S,\text{tr}}}(r_\text{fr},t_\text{fr}) &= n_{S,\text{INCG}}[T, \mu_B, \mu_{I_3}, \mu_S](r_\text{fr},t_\text{fr}), \label{eq:strange}
}
where $r_\text{fr},t_\text{fr}$ corresponds to the coordinate in the center of the volume $(0.5 ~\mathrm{fm})^3$ that is at $n_\text{fr}$  and is used to obtain the event-average $\mean{\cdots}$ of these quantities. 
The energy-momentum tensor in the local rest frame (defined as the collective flow $4$-velocity $u^{\mu}=0$) carries information about pressure and energy density $\varepsilon$, making it connected to the EoS and to thermodynamic properties such as temperature.
The average of $2\times 10^5$ events allowed us to construct a statistically unified ensemble and suppress fluctuations. With temperature and chemical potential at hand, one can use equilibrium distributions to obtain phase-space densities of clusters and nucleons~\cite{Cooper:1974mv}.

Coarse-Grained (CG) descriptions are commonly used in physics to reduce the complexity of the system to the level of detail consistent with practical scales. 
For example, CG is commonly used in polymer physics to simplify molecular chains into beads and springs~\cite{doi1988theory}, in meteorology to resolve large-scale weather patterns from local turbulence~\cite{SMAROGINSKY}, and in effective field theories for particle physics where high-energy modes can be integrated out to access low-energy physics~\cite{WEINBERG1979327}, or in the hydrodynamic description of heavy-ion collisions at ultra-relativistic energies~\cite{Romatschke:2017ejr,Heinz:2024jwu}.

Applying this philosophy to the low-energy HICs at freeze-out, we focus on the late-stage expansion around mid-rapidity, where we expect the approximation of chemical equilibrium to succeed.
The value for the freeze-out density $n_\text{fr}$ is estimated to be $n_\text{fr}\equiv 0.02~\mathrm{fm^{-3}}$ of the order of 1/8 of saturation density. This value agrees with some previous estimates of chemical freeze-out at these energies~\cite{Motornenko:2021nds, Harabasz_2020}. %Although t
The scattering of particles after chemical freeze-out is left for future work
since it
should not significantly alter the chemical composition and will only slightly affect the particle momentum distribution.

\begin{figure}
\centering
\includegraphics[width=0.49\textwidth]{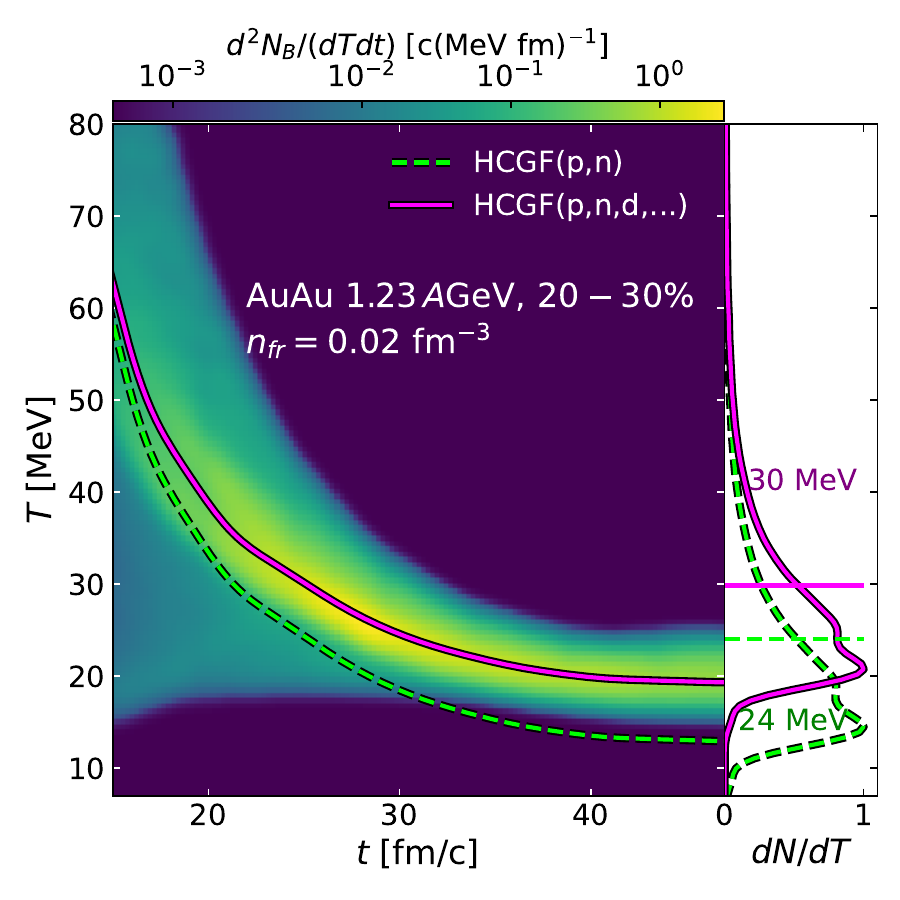}
\caption{\label{fig:hst-001}  Main panel: distribution of emitted baryons associated with the freeze-out density $n_\text{fr}$ in a UrQMD 4.0 transport simulation, as a function of time and temperature calculated using ideal nucleons-and-clusters gas (INCG). The magenta line shows our coarse-grained average temperature, including clusters;  the green line shows the corresponding temperature without clusters. The shift (from green) to higher temperatures (magenta) reflects the energy released as clusters form. The right panel shows the cumulative $T$ distributions for the two models. Horizontal lines show the average $T$. 
}
\end{figure}

The Hybrid CG Freeze-out (HCGF) approach introduced here is agnostic with respect to the transport description or the thermal model, provided that the produced particles and their available states are fixed. Here, we demonstrate this approach with the UrQMD 4.0~\cite{Bass:1998ca, Bleicher:1999xi, Steinheimer:2024eha} transport model and a momentum-dependent mean-field potential calibrated at  to HADES data~\cite{Steinheimer:2025hsr}. These calculations are matched at $n_{B}$=$n_\text{fr}$ via Eqs.~\eqref{eq:energy}-\eqref{eq:strange} to a modified ideal nucleon-and-cluster gas (INCG) EoS that
incorporates corrections to the equation of state due to the presence of hydrogen and helium isotopes at freeze-out (e.g.~\cite{Ren:2023ued} and refs. therein). 
For simplicity, we approximate the proton-neutron density difference $n_{I_3}$ for the system by it's average value 
in the transport model: $n_{I_3}=-0.07 n_B$ ($n_Q=0.43 n_B$)
and set strangeness $n_S=0$~fm$^{-3}$. 
While our framework allows fluctuations in isospin and strangeness densities, these fluctuations are very small for the system of interest. 
In the following, we show that the resulting HCGF calculation provides a reasonable description of both thermal and collective features at chemical freeze-out.

Applying the matching relations~\eqref{eq:energy}-\eqref{eq:strange}, one obtains the distribution of the number of baryons emitted over time and local temperature shown in Fig.~\ref{fig:hst-001} for semi-peripheral ($20\text{--}30\%$ centrality) Au$+$Au collisions at incident energy $E_\text{lab}=1.23~A\mathrm{GeV}$. Collisional mixing of projectile and target nucleons heats the matter. Freeze-out begins in the outer layers and extends into the interior of the fireball over time. The large, hot system initially cools via expansion, decreasing the average temperature over time, as shown by the solid line.  
When a system allows the formation of light clusters, the energy released by cluster formation raises the freeze-out temperature (magenta curve) above that 
for a nucleon-dominated final state (green curve). 
The majority of particles emitted near mid-rapidity typically freeze-out in $t_\text{fr}\approx 30 ~\mathrm{fm/c}$.

\begin{figure}[h!]
\centering
\includegraphics[width=0.49\textwidth]{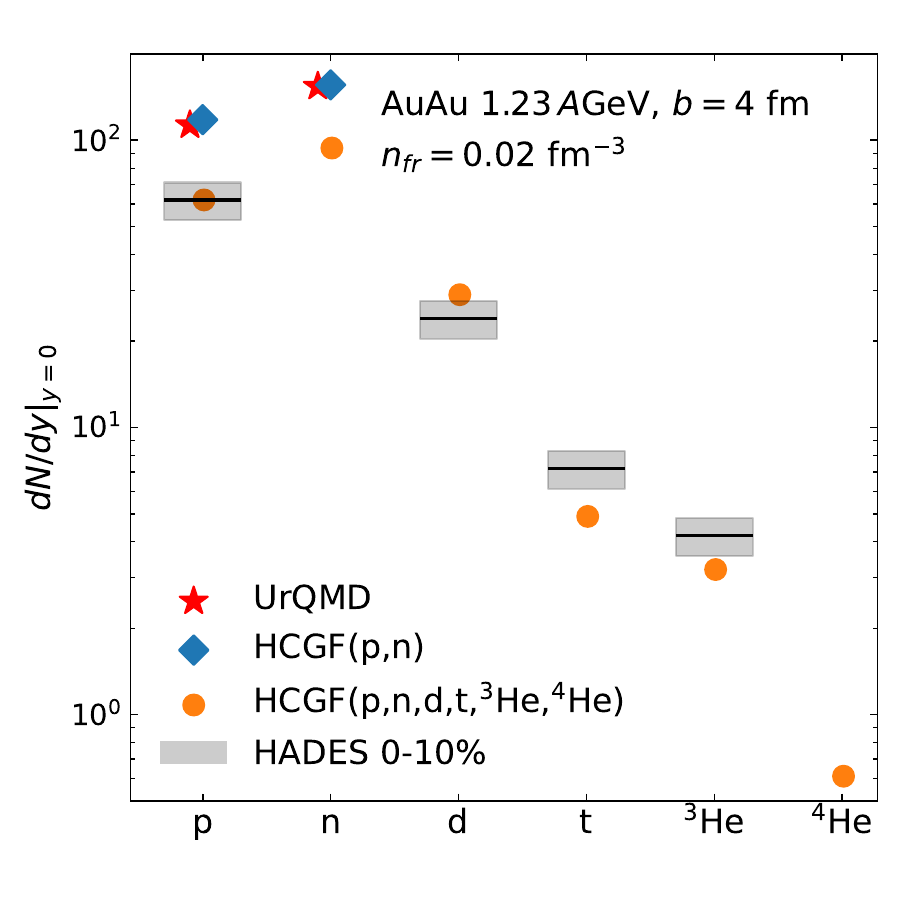}
\caption{\label{fig:yields}Yields of protons, neutrons, and light clusters from HCGF(diamonds and circles) compared to the original UrQMD transport model (stars) and the HADES experimental data (grey band).} 
\end{figure}

The influence of cluster formation on the temperature can be illustrated by comparing two simple models:
1) with only nucleons $N$ and temperature $T$, and 2) with nucleons $N$ and deuterons $d$ and temperature $T^\prime$; the latter calculation illustrates the coalescence heating due to clusterization. 
We assume Boltzmann distributions and a non-relativistic system.
There are three translational degrees of freedom per gas nucleus, so the energy of the nucleonic system is $E= (m + \frac 3 2 T)n_N$, and the energy of the nucleon+deuteron system is $E^\prime=m(n_N^\prime+2n_d^\prime)+\frac 3 2 T^\prime(n_N^\prime+n_d^\prime)-B n_d^\prime$ ($d$ are point particles same as $N$), where $m$ is the bare mass. The 2.2 MeV release in binding energy will heat the system in the case of deuteron formation, but the temperature would nevertheless be increased even if the binding energy is neglected. 
Assuming the same matching condition for the energy and baryon number, see Eqs.~\eqref{eq:energy} and \eqref{eq:nb},
the following relation between the temperatures 
can be obtained:
\begin{equation}
T^\prime = \frac{2}{1+\lambda} T \, ,
\end{equation}
where $\lambda=n_N^\prime/n_N$ represents the fraction of nucleons outside of clusters. Since $\lambda\in [0,1]$, see Fig.~\ref{fig:yields} for instance, we have $T^\prime \geq T$, in qualitative agreement with our results shown in Fig.~\ref{fig:hst-001}. 

\begin{figure}[h]
\centering
\includegraphics[width=0.49\textwidth]{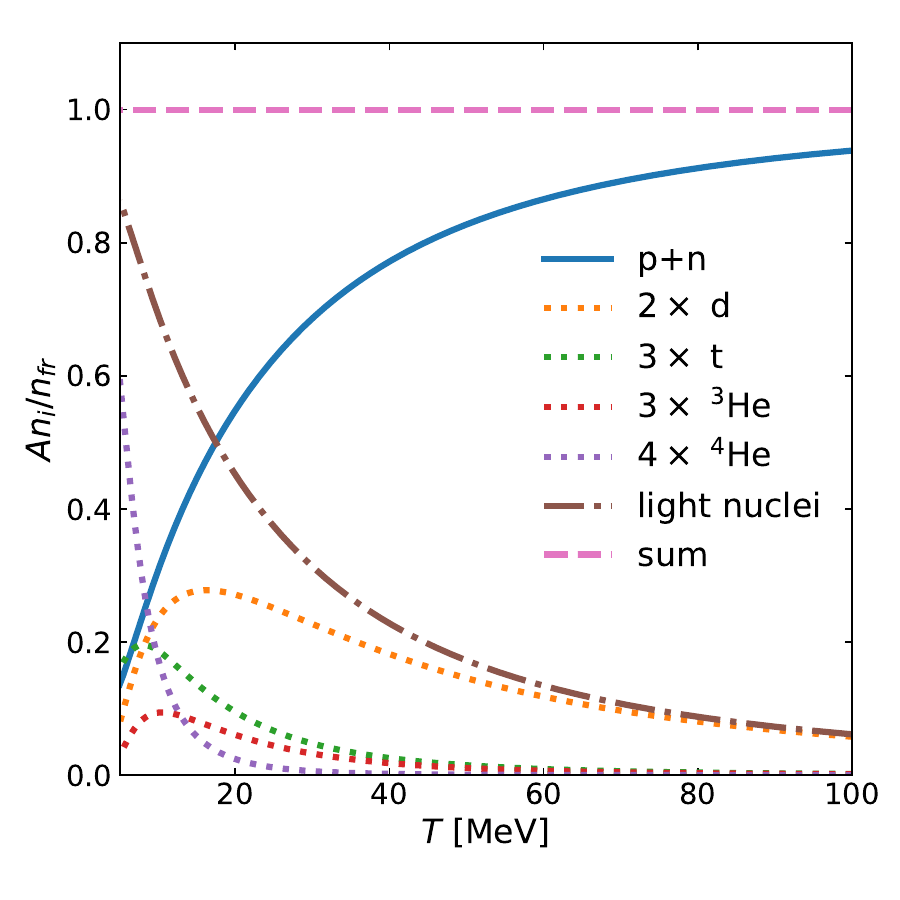}
\caption{\label{fig:yields-vs-T} The yields of nucleons and light fragments as a function of temperature under the constraint of fixed total baryon and charge densities. At high temperature, entropy favors a high number of degrees of freedom and few clusters. This is contrasted with a depletion of the free nucleon fraction in favor of lighter clusters at low temperatures due to the shrinkage of available phase space. At the lowest temperature scales, $\alpha$-particles are predicted to be the dominant surviving species due to their high binding energy and low degeneracy pressure compared to free nucleons. }
\end{figure}

On the right side of Fig.~\ref{fig:hst-001}, the distributions of temperature marginalized over time are shown for the HCGF(p,n) and HCGF(p,n,d,t,$^3$He,$^4$He) models. Their centroids are respectively located at 24~MeV and 30~MeV,
confirming that the model with clusters is warmer than the model with only nucleons.

The relation between the freeze-out temperature and the density is expected to affect cluster yields. Figure~\ref{fig:yields-vs-T} shows a relative fraction of nucleon and clusters at the fixed baryon density $n_\text{fr}$ and electric charge $n_Q=0.43n_\text{fr}$ as a function of temperature. Of course, due to the binding energy, the phase space available to light clusters is effectively larger, strongly favoring their production at low temperatures, where phase space is otherwise restricted. Conversely, at high temperatures, the drive toward maximum entropy favors a higher number of degrees of freedom; thus, a state of two free nucleons is statistically preferred over one deuteron. This picture must, however, be amended to account for in-medium reactions, where protons and neutrons form light clusters due to dynamic correlations. These correlations introduce an additional constraint, namely chemical equilibrium, which maintains a small fraction of light clusters even at high temperatures, though they remain strongly suppressed relative to free nucleons. This also implies that the ratio of proton to light-cluster emission should not be constant. Instead, free protons will predominantly come from earlier stages, while light clusters will come from later ones.

\begin{figure}[h!]
\centering
\includegraphics[width=0.49\textwidth]{
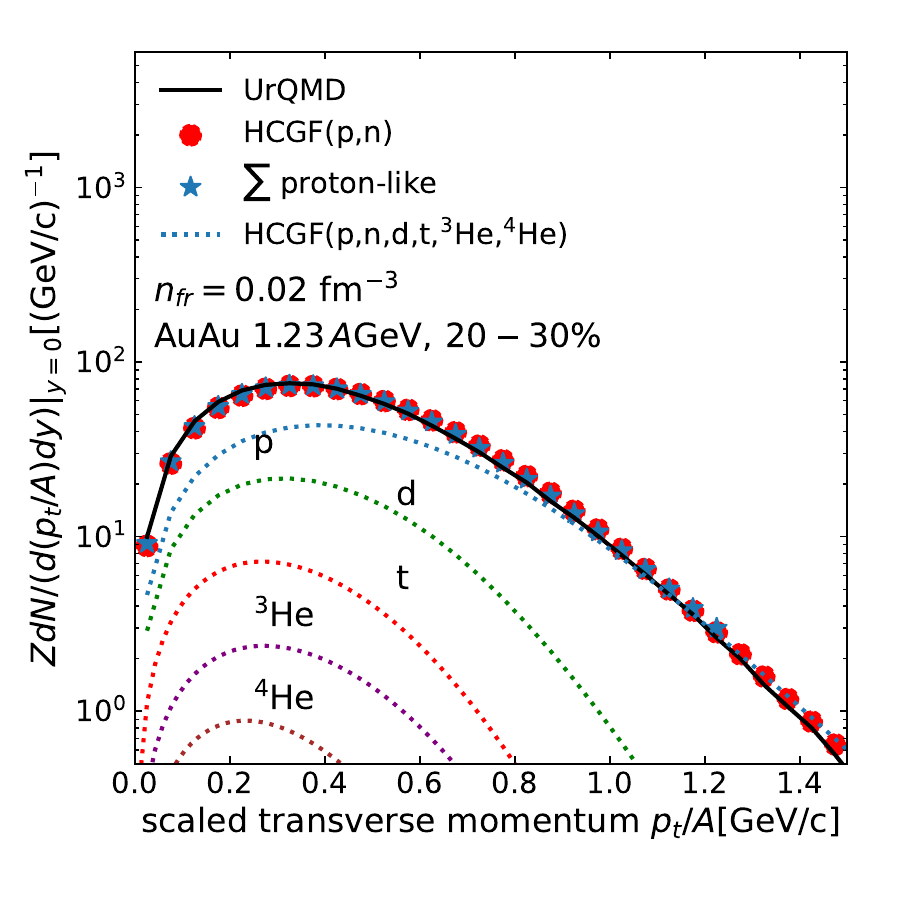}
\caption{\label{fig:spectra} Scaled spectra for charged particles at mid-rapidity in different model versions as a function of the transverse momentum scaled with mass number. Model versions include UrQMD (solid line), HCGF(p,n) (red circles), and HCGF(p,n,d,t,$^3$He,$^4$He) (blue stars). In the last case, a proton-like spectrum (dotted lines) is also shown.}
\end{figure}

We now illustrate the capacity of the HCGF approach to 
predict different observables, 
and we start with the more global one: particle yields. The stars in Fig.~\ref{fig:yields} represent the proton and neutron yields obtained from UrQMD transport simulation without coalescence. It is compared with the proton and neutron yields obtained from HCGF(n,p) model (blue diamond).
We also show the yields for protons, neutrons, and light clusters (yellow circles) in Fig.~\ref{fig:yields}. 
For simplicity, we consider only ground state energies here, but the inclusion of excited states and their decays may, in some cases, be necessary to accurately describe the observed yields~\cite{Vovchenko:2020dmv}. The extension to those more complex processes~\cite{Blaschke:2024jqd, Rischke:1991ke} is relatively straightforward and will be implemented in the future. 
Fig.~\ref{fig:yields} shows that the HCGF model reproduces very well the experimental data from HADES~\cite{Szala:2020jro} (grey band).

We now analyze particle spectra as a function of the scaled transverse momentum for protons and light clusters. The proton spectrum is shown in Fig.~\ref{fig:spectra} for UrQMD without any explicit clustering (solid black line) and for the HCGF(p,n) approach (red circles). There is a good overlap between these two spectra.
The other curves show the spectra for protons and light clusters obtained from the HCGF(p,n,d,t,$^3$He,$^4$He) approach. 
Light cluster spectra are peaked at lower scaled transverse momentum than those of protons. The asymptotic slope of the spectra reflects the effective temperature of the emission region, which is related to its temperature and, more importantly, to its collective flow~\cite{PhysRevLett.42.880,Gaitanos:1999qr}. Since the collective flow dominates here over temperature, the spectra do not provide conclusive information about the sources' temperatures. At the same time, clusters predominantly populate states at low scaled transverse momentum, consistent with their late emission when transverse flow and temperature are expected to be smaller according to common blast-wave approaches.

The spectra shown in Fig.~\ref{fig:spectra} were obtained after integrating the local distribution of nucleons and clusters at the temperature and chemical potentials obtained in the HCGF approach. 
Spectra contain information regarding the total charge — where the integral over the spectra equals the net charge — and 
the total energy, 
where the integral of energy ($E=\sqrt{p^2_t+m_i^2}\cosh y$) over the spectra represents the total energy of the system with $m_i$ the mass of the particle (including the binding energy).
Thus, the energy of protons and neutrons obtained from UrQMD matches the energy of the protons, neutrons, and light clusters from the HCGF(p,n,d,t,$^3$He,$^4$He) approach.

\begin{figure}
\centering
\includegraphics[width=0.49\textwidth]{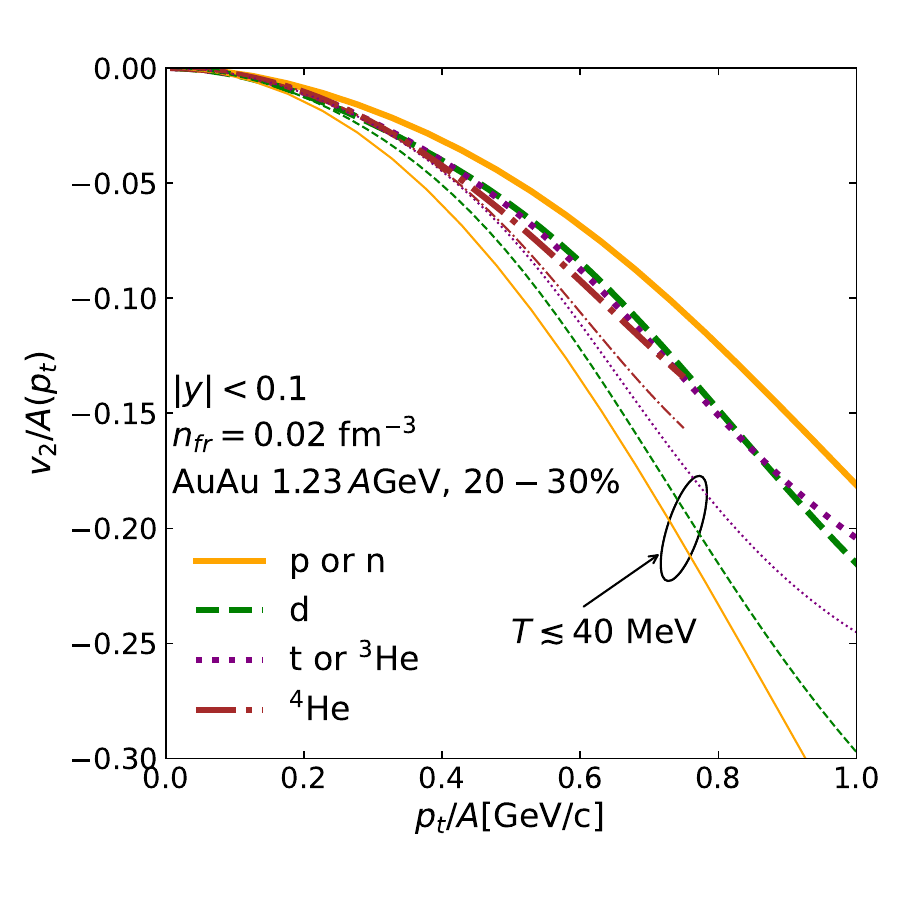}
\caption{\label{fig:v2-pt} Thick lines show elliptic flow $v_{2}/A$ of protons and light fragments emitted from the entire freeze-out in the HCGF(p,n,d,t,$^3$He,$^4$He) model as a function of the transverse momentum $p_t/A$ scaled with mass number $A$. Thin lines represent results obtained using contributions from a subset of the freeze-out hypersurface with $T\lesssim 40$~MeV.}
\end{figure}

Finally, we analyze the elliptic flow as a function of the transverse momentum for protons, neutrons, and light clusters within the HCGF(p,n,d,t,$^3$He,$^4$He) approach. We show in Fig.~\ref{fig:v2-pt} the scaled elliptic flow as a function of the scaled transverse momentum. The thick lines are obtained by considering the full temperature distribution, whereas the thin lines represent a sensitivity analysis in which emission from regions with temperatures above 40~MeV has been excluded. The flow of protons is clearly more impacted, demonstrating that nucleons come from regions of the system that have a wider range of freeze-out temperatures than clusters.
At low transverse momentum $p_t/A\lesssim 0.2$~GeV/c, there is a good overlap of the elliptic flow for the different particles considered in our work, and only a small effect of the distribution of large temperatures, $T\gtrsim 40$~MeV. Differences between protons and light clusters appear at larger transverse momenta ($p_t/A\gtrsim 0.4$~GeV/c), where the effect of $T$ is also larger.

In summary, we propose the hybrid coarse-grained freeze-out model HCGF(p,n,d,t,$^3$He,$^4$He) for mid-rapidity particles and nuclear clusters. This model describes the final stage of HIC in which some protons and neutrons combine to form clusters. The model receives input from transport simulations averaged over a large ensemble, from which local sources are extracted at different temperatures and collective velocities, yielding light-particle distributions that can be directly compared with data. With different temperatures and collective velocities for the sources, our model can address yields, spectral shapes, and flows of protons, neutrons, and light clusters.

The HCGF framework provides a thermodynamically consistent method for cluster production at mid-rapidity that preserves the dynamical features of transport models and captures the internal energy of bound states. Several corrections can be made in the future: high-$p_t$ particles can be treated separately, and corrections to out-of-equilibrium processes can be included within the DNMR~\cite{Denicol:2012cn} or Chapman-Enskog~\cite{chapman1970mathematical} expansions, i.e., by modifying the distribution functions used in calculations. 
Additional improvements to the model should also be considered to facilitate comparison with the data, including excited states and decays, as well as particles at large rapidity.

\section*{Acknowledgments}

OS and JM both benefit from the IN2P3 Masterproject MAC2.
This work was supported by the U.S.\ Department of Energy, Office of Science, under Grant No.\ DE-SC0019209. 

\bibliography{refs}

\end{document}